\documentstyle[11pt,epsf]{article}
\newtheorem{theorem}{Theorem}

\newcommand {\dfn} {\stackrel{\Delta} {=}}

\newcommand {\bb} {\mbox{\boldmath $b$}}

\newcommand {\bk} {\mbox{\boldmath $k$}}
\newcommand {\bl} {\mbox{\boldmath $l$}}

\newcommand {\bs} {\mbox{\boldmath $s$}}

\newcommand {\bu} {\mbox{\boldmath $u$}}

\newcommand {\bx} {\mbox{\boldmath $x$}}
\newcommand {\by} {\mbox{\boldmath $y$}}
\newcommand {\bz} {\mbox{\boldmath $z$}}

\newcommand {\bE} {\mbox{\boldmath $E$}}

\newcommand {\hX} {\hat{X}}
\newcommand {\hY} {\hat{Y}}
\newcommand {\hy} {\hat{y}}
\newcommand {\hx} {\hat{x}}

\newcommand{\calA}{{\cal A}}
\newcommand{\calB}{{\cal B}}

\newcommand{\calE}{{\cal E}}

\newcommand{\calK}{{\cal K}}

\newcommand{\calS}{{\cal S}}

\newcommand{\calX}{{\cal X}}
\newcommand{\calY}{{\cal Y}}
\newcommand{\calZ}{{\cal Z}}

\begin{document}
\thispagestyle{empty}
\title{Perfectly Secure Encryption of Individual Sequences
\thanks{This research was supported by ISF grant no.\ 208/08.}
}
\author{Neri Merhav
}
\date{}
\maketitle

\begin{center}
Department of Electrical Engineering \\
Technion - Israel Institute of Technology \\
Technion City, Haifa 32000, ISRAEL \\
E--mail: {\tt merhav@ee.technion.ac.il}\\
\end{center}
\vspace{1.5\baselineskip}
\setlength{\baselineskip}{1.5\baselineskip}

\begin{abstract}
In analogy to the well--known notion of finite--state compressibility of individual
sequences, due to Lempel and Ziv, we define a similar notion of ``finite--state
encryptability'' of an individual plain-text sequence, as the minimum asymptotic key rate
that must be consumed by finite--state encrypters so as to guarantee perfect secrecy in a
well--defined sense. Our main basic result is that the finite--state encryptability
is equal to the finite--state compressibility for every individual sequence. 
This is in parallelism to Shannon's classical
probabilistic counterpart result, asserting that the minimum required key rate is equal
to the entropy rate of the source. However, the redundancy, defined
as the gap between
the upper bound (direct part) and the lower bound (converse part) in the
encryption problem, turns out to decay at a different rate (in fact, much slower) than the
analogous redundancy associated with the compression problem. We also extend
our main theorem in several directions, allowing: (i) availability of side
information (SI) at the encrypter/decrypter/eavesdropper, (ii) lossy
reconstruction at the decrypter, and (iii) the combination of both lossy
reconstruction and SI, in the spirit of the Wyner--Ziv problem.
\\

\end{abstract}

\section{Introduction}

The paradigm of individual sequences and finite--state machines (FSMs), as an
alternative to the traditional probabilistic modeling of sources and channels,
has been studied and explored quite extensively in several 
information--theoretic problem areas,
including data
compression \cite{KY}, \cite{MZ06}, \cite{RM11}, \cite{WMF94}, \cite{WM02},
\cite{Ziv78}, \cite{Ziv80}, \cite{Ziv84}, \cite{ZL78}, source/channel
simulation \cite{MMSW10}, \cite{Seroussi06}, classification \cite{Ziv88}, \cite{ZM93}, 
prediction \cite{FMG92}, \cite{HKW98}, \cite{MF93}
\cite{WM01}, \cite{WMSB01}, 
\cite{ZM07}, denoising \cite{WOSVW05}, 
and even channel coding \cite{LF10}, \cite{SF05},
just to name very few representative references out of many more.
On the other hand, it is fairly safe to say that the entire literature on 
information---theoretic security, starting from Shannon's
classical work \cite{Shannon48} and ending with some of the most recent work in this
problem area (see, e.g., \cite{Hellman77},
\cite{Lempel79}, \cite{LPS09}, \cite{Massey88}, \cite{Yamamoto91} 
for surveys as well as references therein), is based
exclusively on the probabilistic setting. 

To the best of our knowledge, the only exception to
this rule is an unpublished memorandum by Ziv \cite{Ziv78u}.
In that work, the plain-text source to be encrypted, using a secret key, is an
individual sequence,
the encrypter is a general block encoder, and the eavesdropper
employs an FSM as a message discriminator. Specifically, it is postulated in
\cite{Ziv78u} that
the eavesdropper may have some prior knowledge about the plain-text that can be expressed in terms of
the existence of some set of ``acceptable messages'' that constitutes the
a-priori level of uncertainty (or equivocation) that the eavesdropper has concerning the
plain-text message: The larger the acceptance set, the larger is the
uncertainty. Next, it is assumed that there exists
an FSM that can test whether a given
candidate plain-text message is acceptable or not: If and only if the FSM produces the all--zero
sequence in response to that message, then this message is acceptable.
Perfect security is then defined as a situation where the
size of the acceptance set is not reduced (and hence neither is the
uncertainty) in the presence of the cryptogram.
The main result in \cite{Ziv78u} is that the asymptotic key rate needed for
perfectly secure encryption in that sense, cannot be smaller (up to
asymptotically vanishing terms) than the
Lempel--Ziv (LZ) complexity of the plain-text source \cite{ZL78}. This lower bound
is obviously asymptotically achieved by one--time pad encryption of the bit-stream obtained by
LZ data compression of the plain-text source.
This is in parallelism to Shannon's classical
probabilistic counterpart result, asserting that the minimum required key rate is equal
to the entropy rate of the source.

In this paper, we also consider encryption of individual sequences, but
our modeling approach and the definition of perfect secrecy are 
substantially different. Rather than assuming
that the encrypter and decrypter have unlimited resources, 
and that it is the eavesdropper
which has limited resources, modeled in terms of FSMs, in our setting, the
converse is true. We adopt a model of 
a finite--state encrypter, which receives as
inputs the plain-text stream and the secret key bit-stream, and it produces a
cipher-text, while the internal state variable of the FSM, that designates limited memory of the past
plain-text, is evolving in response to the plain-text input.
Based on this model, we define a
notion of {\it finite--state encryptability} (in analogy to the
notions of finite--state compressibility 
\cite{ZL78} and the finite--state predictability \cite{FMG92}), 
as the minimum
achievable rate at which 
key bits must be consumed by any finite--state
encrypter in order to guarantee
perfect security against an unauthorized party,
while keeping the cryptogram
decipherable at the legitimate receiver, which has access to the key.
Our main result is that the finite--state encryptability is equal to the
finite--state compressibility, similarly as in \cite{Ziv78u}. 

More precisely,
denoting by $c(x^n)$ the number of LZ phrases associated with the
plain-text $x^n=(x_1,\ldots,x_n)$, we show that number of key bits required
by any encrypter with $s$ states,
normalized by $n$ (i.e., the key rate), cannot be smaller than $[c(x^n)\log
c(x^n)]/n-\delta_s(n)$, where $\delta_s(n)=O(s\log(\log n)/\sqrt{\log n})$.
On the other hand, this bound is obviously essentially achievable by applying
the LZ `78 algorithm \cite{ZL78}, followed by one--time pad encryption (i.e.,
bit--by--bit XORing between compressed bits and key bits), since the
compression ratio of the LZ `78 algorithm is also
$[c(x^n)\log c(x^n)]/n$, up to vanishingly small terms. It follows then that
the finite--state encryptability of every (infinite) individual sequence is equal to its
finite--state compressibility. 

While the idea of LZ data compression,
followed by one--time padding is rather straightforward, our main result, that
no finite--state encrypter can do better than that for any given individual
sequence, may not be quite obvious since
the operations of compression and encryption are basically different --
secret key encryption need not necessarily be 
based on compression followed by one--time
padding, definitely not if both operations are 
formalized in the framework of finite--state machines.

For finite sequences of length $n$, the difference between the upper
bound (of the direct part) and the lower bound
(of the converse part), which can be thought of as some notion of
redundancy, is again $O(s\log(\log n)/\sqrt{\log n})$, which
decays much more slowly than the corresponding redundancy in data compression
\cite[Theorems 1, 2]{ZL78}, which is roughly
$O((\log s)/\log n)$. 

Finally, we extend our main basic theorem in two directions, first, one at a
time, and then simultaneously.
The first extension is in allowing availability of side information (SI) at all three parties
(encrypter, legitimate decrypter and eavesdropper) or at the decrypter and the
eavesdropper only. We assume that the SI sequence is an
individual sequence as well. We also assume that it is the same SI
that is available to all three parties in the first case or to
both the legitimate decrypter and the eavesdropper, in the second case. 
Extensions to situations of different versions of the SI at
different users is deferred to the last step, which will possess the most
general scenario we study in this work.
Our main result is essentially unaltered, except that the LZ complexity,
$\rho_{LZ}(x^n)\dfn [c(x^n)\log c(x^n)]/n$, is replaced by the conditional LZ
complexity given the SI, to be defined later (see also \cite{Merhav00}, \cite{Ziv85}).
Our second extension is to the case where lossy reconstruction is allowed at
the legitimate receiver (first, without SI). Here the LZ complexity is
replaced by a notion of ``LZ rate--distortion 
function,'' $r_{LZ}(D;x^n)$, which means the smallest
LZ complexity among all sequences 
that are within the allowed distortion relative to
the input plain-text sequence. While our framework allows randomized reconstruction
sequences (that may depend on the random key), we find that at least
asymptotically, there is nothing to gain from this degree of freedom, as
optimum performance can be achieved by a scheme that generates deterministic
reproductions. Finally, we allow both SI and lossy reconstruction at the same
time. Moreover, every party might have access to a different version of the SI.
The SI available to the legitimate receiver is assumed to be generated by the
plain-text source via a known memoryless channel.
Here we no longer characterize the performance in terms  
LZ complexities of sequences, but rather in the same spirit of the Wyner--Ziv rate--distortion
function for individual sequences using finite--state encoders and decoders
\cite{MZ06}. 

It should be pointed out that throughout the entire paper, most
of our emphasis is on converse theorems (lower bounds). The compatible direct parts
(upper bounds) will always be attainable by a straightforward application of the
suitable data compression scheme, followed by one--time padding.

The outline of the remaining part of this paper is as follows.
In Section 2, we establish some notation conventions and we formally define
the model and the problem. In Section 3, we assert and prove the main result.
Finally, in Section 4, we extend our results in the above--mentioned
directions, and we point out how exactly the proof of the basic theorem should
be modified in each case in order to support our assertions.

\section{Notation Conventions and Problem Formulation}

We begin by establishing some notation conventions.
Throughout this paper, scalar random
variables (RV's) will be denoted by capital
letters, their sample values will be denoted by
the respective lower case letters, and their alphabets will be denoted
by the respective calligraphic letters.
A similar convention will apply to
random vectors and their sample values,
which will be denoted with same symbols superscripted by the dimension.
Thus, for example, $A^m$ ($m$ -- positive integer)
will denote a random $m$-vector $(A_1,...,A_m)$,
and $a^m=(a_1,...,a_m)$ is a specific vector value in $\calA^m$,
the $m$--th Cartesian power of $\calA$. The
notations $a_i^j$ and $A_i^j$, where $i$
and $j$ are integers and $i\le j$, will designate segments $(a_i,\ldots,a_j)$
and $(A_i,\ldots,A_j)$, respectively,
where for $i=1$, the subscript will be omitted (as above).
For $i > j$, $a_i^j$ (or $A_i^j$) will be understood as the null string.

Sources and channels will be denoted generically by the letter $P$ or $Q$, 
subscripted by the name of the RV and its conditioning,
if applicable, exactly like in
ordinary textbook notation standards, e.g., $P_{X^m}(x^m)$ is the probability function of
$X^m$ at the point $X^m=x^m$, $P_{W|S^m}(w|s^m)$
is the conditional probability of $W=w$ given $S^m=s^m$, and so on.
Whenever clear from the context, these subscripts will be omitted.
Information theoretic quantities, like entropies and mutual
informations, will be denoted following the usual conventions
of the information theory literature, e.g., $H(K^m)$, $I(W;X^m|S^m)$,
and so on.

A finite--state encrypter is defined by a sixtuplet
$E=(\calX,\calY,\calZ,f,g,\Delta)$,
where $\calX$ is a finite input alphabet of size $|\calX|=\alpha$,
$\calY$ is a finite set of binary words,
$\calZ$ is a finite set of states, 
$f:\calZ\times\calX\times\{0,1\}^*\to\calY$ is the
output function, $g:\calZ\times\calX\to\calZ$ is the
next--state function, 
$\Delta:\calZ\times\calX\to \{0,1,2,\ldots\}$, and $\{0,1\}^*$ is the
set of all binary strings of finite length.
The set $\calY$ is allowed to contain binary
strings of various lengths, including the null word $\lambda$ (whose length is zero).
When two infinite sequences, $\bx=x_1,x_2,\ldots$, $x_i\in\calX$, henceforth
the {\it plain-text sequence} (or, the source sequence),
and $\bu=u_1,u_2,\ldots$, $u_i\in\{0,1\}$, 
$i=1,2,\ldots$, henceforth the {\it key sequence}, 
are fed into an encrypter $E$, it produces an infinite output sequence 
$\by=y_1,y_2,\ldots$, $y_i\in\calY$, henceforth the {\it cryptogram}, 
while passing through an infinite
sequence of states 
$\bz=z_1,z_2,\ldots$, $z_i\in\calZ$, according to
the following recursive equations, implemented for $i=1,2,\ldots$
\begin{eqnarray}
t_i&=&t_{i-1}+\Delta(z_i,x_i),~~~~~~t_0\dfn 0 \label{ti}\\
k_i&=&(u_{t_{i-1}+1},u_{t_{i-1}+2},\ldots,u_{t_i}) \label{ki}\\
y_i&=&f(z_i,x_i,k_i) \label{yi}\\
z_{i+1}&=&g(z_i,x_i) \label{nextstate}
\end{eqnarray}
where it is understood that if 
$\Delta(z_i,x_i)=0$, then $k_i=\lambda$, the null word of
length zero,\footnote{Note that the evolution of the state $z_i$ depends only
on the source inputs $\{x_i\}$, not on the key bits. The rationale is that the
role of $z_i$ is to store past memory of the information sequence $x^n$, in order to
take advantage of empirical 
correlations and repetitive patterns 
in that sequence, whereas memory of past key
bits, which are i.i.d., is irrelevant. 
Nonetheless, it is possible to extend the encrypter model to have
two separate state variables, one evolving with dependence on $\{x_i\}$ only
(as above) and one
with dependence on both $\{x_i\}$ and $\{k_i\}$, where the former state
variable plays a role in the update of $t_i$ 
and the latter plays a role in the output
function.} namely, no key bits are used in the $i$--th step.
By the same token, if $y_i=\lambda$, no output is produced at this
step, i.e., the system is idling and only the state evolves in response
to the input.
An encrypter with $s$ states, or an $s$--state encrypter,
$E$, is one with $|\calZ|=s$.
It is assumed that the plain-text sequence $\bx$ is deterministic (i.e., an
individual sequence), whereas 
the key sequence $\bu$ is purely random, i.e.,
for every positive integer
$n$, $P_{U^n}(u^n)=2^{-n}$.

A few additional notation conventions will be convenient: By $f(z_1,x^n,k^n)$, we refer
to the vector $y^n$ produced by $E$ in response to the inputs $x^n$ and $k^n$
when the initial state is $z_1$. Similarly, the notation $g(z_1,x^n)$ will mean the
state $z_{n+1}$ and $\Delta(z_1,x^n)$ will designate
$\sum_{i=1}^n\Delta(z_i,x_i)$ under the same circumstances. 
An encrypter $E$ is said to be
{\it perfectly secure} if for every two positive integers $n$, $m$
($m\ge n$) and for every $\bx\in\calX^{\infty}$ and
$y_n^m\in\calY^{m-n+1}$, the probability
$\mbox{Pr}\{Y_n^m=y_n^m|\bx\}$ is independent of $\bx$.

An encrypter is referred to as {\it information lossless} (IL) if for every
$z_1\in\calZ$, every sufficiently large\footnote{
It should be pointed out that this definition of information losslessness
is more relaxed (and hence more general) than the definition in \cite{ZL78}. While
in \cite{ZL78}, the requirement is imposed for {\it every} positive integer $n$, here
it is required only for all sufficiently large $n$. Note that lack of information
losslessness in the more restrictive sense of \cite{ZL78} is not in contradiction
with the ability to reconstruct the source at the legitimate decoder. All it
means is that reconstruction of $x^n$ may require more information than just
$(z_1,y^n,k^n,z_{n+1})$, for example, 
some additional data from times later than
$n+1$ may be needed.}
$n$ and all 
$x^n\in\calX^n$ and $k^n\in \calK^n$, the quadruple
$(z_1,k^n,f(z_1,x^n,k^n),g(z_1,x^n))$ uniquely determines $x^n$. 
It will henceforth be assumed,
without loss of generality,
that $z_1$ is a certain fixed member of $\calZ$.
Given an encrypter $E$ and an input string $x^n$, the encryption key rate of
$x^n$ w.r.t.\ $E$ is defined as
\begin{equation}
\sigma_E(x^n)\dfn \frac{\ell(k^n)}{n}=\frac{1}{n}\sum_{i=1}^n\ell(k_i),
\end{equation}
where $\ell(k_i)=\Delta(z_i,x_i)$ is the length of the binary string $k_i$ and
$\ell(k^n)=\sum_{i=1}^n\ell(k_i)$ is the total length of $k^n$.

The set of all perfectly secure, IL encrypters $\{E\}$ with no more than $s$
states will be denoted by $\calE(s)$.
The minimum of $\sigma_E(x^n)$ over all encrypters in $\calE(s)$
will be denoted by $\sigma_s(x^n)$, i.e.,
\begin{equation}
\sigma_s(x^n)=\min_{E\in \calE(s)}\sigma_E(x^n).
\end{equation}
Finally, let 
\begin{equation}
\sigma_s(\bx)=\limsup_{n\to\infty}\sigma_s(x^n),
\end{equation}
and define the {\it finite--state encryptability} of $\bx$ as
\begin{equation}
\sigma(\bx)=\lim_{s\to\infty}\sigma_s(\bx).
\end{equation}
Our purpose it to characterize these quantities and to point out
how they can be achieved in principle.

\section{Main Result}

Incremental parsing \cite{ZL78} of a string $x^n$ is a sequential procedure of
parsing $x^n$ into distinct phrases, where each new parsed phrase is the
shortest string that has not been encountered before as a phrase of $x^n$,
with the possible exception of the last phrase that might be incomplete.
Let $c(x^n)$ denote the number of phrases in LZ incremental parsing of $x^n$.
The LZ complexity of $x^n$ is defined as
\begin{equation}
\rho_{LZ}(x^n)\dfn \frac{c(x^n)\log c(x^n)}{n}.
\end{equation}
The finite--state compressibility, $\rho(\bx)$, of the infinite sequence
$\bx=(x_1,x_2,\ldots)$ is defined, in \cite{ZL78}, as the best
compression ratio achieved by IL finite--state encoders,
analogously to the above definition of finite--state encryptability.
From Theorems 1, 2 and 3 of \cite{ZL78},
it follows that $\rho_{LZ}(\bx)\dfn\limsup_{n\to\infty}\rho_{LZ}(x^n)$ 
is equal to $\rho(\bx)$.

The following theorem establishes a lower bound on $\sigma_s(x^n)$ in terms of
$\rho_{LZ}(\bx^n)$ and hence a lower bound
of $\sigma(\bx)$ in terms of $\rho(\bx)$.
\begin{theorem}
(Converse to a coding theorem): For every $x^n$,
\begin{equation}
\sigma_s(x^n)\ge \rho_{LZ}(x^n)-\delta_s(n),
\end{equation}
where $\delta_s(n)$ is independent of $x^n$ and behaves according to
\begin{equation}
\delta_s(n)=O\left(\frac{s\log(\log n)}{\sqrt{\log n}}\right).
\end{equation}
Consequently, $\sigma(\bx)\ge\rho(\bx)$.
\end{theorem}

\noindent
{\it Discussion.} A few comments on Theorem 1 are in order.
\begin{enumerate}
\item
It is readily observed that a compatible direct theorem holds, simply by
applying the LZ `78 algorithm followed by one--time pad encryption of the compressed
bits. The resulting key--rate needed is then upper bounded by 
$\frac{1}{n}[c(x^n)+1]\log[2\alpha(c(x^n)+1)]$, following \cite[Theorem
2]{ZL78},
which is, within negligible terms, equal to $\rho_{LZ}(x^n)$.
Thus, $\sigma(\bx)=\rho(\bx)$.
\item Consider the difference between the upper bound pertaining to the
direct part (as mentioned in item
no.\ 1 above) and the lower bound of the converse part. The behavior 
of this difference is 
$O(\alpha s\log(\log n)/\sqrt{\log n})$. This behavior is different from the 
behavior of the corresponding gap in compression (Theorems 1 and 2 in
\cite{ZL78}), which is
$O([\log(2\alpha)]\log(8\alpha s^2)/\log n)$. The guaranteed convergence to
optimality is therefore considerably slower in the encryption problem.
\item As will be seen in the proof of Theorem 1, $\sigma_s(x^n)$ is first
lower bounded in terms of the $m$--th order empirical entropy associated with $x^n$
(namely, the entropy associated with the
relative frequency of non--overlapping $m$--blocks of $x^n$),
where $m$ is a large positive integer,
and then this empirical entropy in turn is further lower bounded in terms of
$\rho_{LZ}(x^n)$. The reason for the latter passage is to get rid of the
dependence of the main term of the lower bound on the parameter $m$, which is
arbitrary. This also helps to select the optimum growth rate of $m$ as a function
of $n$.
\item We already mentioned that the definition of the IL property here is somewhat more
relaxed than in \cite{ZL78} (see footnote no.\ 2). 
Moreover, it is possible to relax this
requirement even further by allowing a relatively small uncertainty in $x^n$ given
$(z_1,k^n,f(z_1,x^n,k^n),g(z_1,x^n))$ (see Subsection \ref{lossy}), at the possible cost of
further slowing down the convergence of $\delta_s(n)$.
\end{enumerate}

\noindent
{\it Proof.} Let $m$ divide $n$ and consider the partition of $x^n$ into
$n/m$ non--overlapping $m$--vectors 
$\bx_1,\bx_2,\ldots,\bx_{n/m}$, where
$\bx_i=x_{(i-1)m+1}^{im}$. 
Recall that for a given $z_{(i-1)m+1}$ and
$\bx_i$, the length $\bl_i$ of 
$\bk_i=k_{(i-1)m+1}^{im}$ is uniquely determined as 
$\bl_i=
\Delta(z_i,x_{(i-1)m+1}^{im})$.
Let us now define a joint empirical distribution of several variables.
For every 
$a^{m}\in\calX^{m}$, $z,z'\in\calZ$, and every positive integer $l$,
let 
\begin{equation}
P_{X^{m}ZZ'L}(a^{m},z,z',l)=\frac{m}{n}\sum_{i=1}^{n/m}
1\{x_{(i-1)m+1}^{im}=a^{m},
z_{(i-1)m+1}=z,z_{im+1}=z',\Delta(z,a^{m})=l\}.
\end{equation}
Now, define
\begin{equation}
P_{K^{m}X^{m}Y^{m}ZZ'L}(\kappa^{m},a^{m},b^{m},z,z',l)=
2^{-l}P_{X^{m}ZZ'L}(a^{m},z,z',l)\cdot
1\{b^{m}=f(z,a^{m},\kappa^{m})\}
\end{equation}
Throughout this proof, all information measures are defined w.r.t.\ 
$P_{K^{m}X^{m}Y^{m}ZZ'L}$. 
Consider the following chain of
equalities for the given $x^n$ and an arbitrary encrypter $E\in\calE(s)$:
\begin{eqnarray}
\sigma_E(x^n)&=&\frac{\ell(k^n)}{n}\nonumber\\
&=&\frac{1}{m}\cdot\frac{m}{n}
\sum_{i=1}^{n/m}\ell(k_{(i-1)m+1}^{im})\nonumber\\
&=&\frac{1}{m}\cdot\frac{m}{n}
\sum_{i=1}^{n/m}H(K_{(i-1)m+1}^{im})\nonumber\\
&=&\frac{H(K^{m}|L)}{m}.
\end{eqnarray}
Note that the length of the key for the $i$-th $m$--block,
$\bl_i=\ell(\bk_i)=\Delta(z_{(i-1)m+1},x_{(i-1)m+1}^{im})=
\sum_{t=(i-1)m+1}^{im}\Delta(z_t,x_t)$, is a variable
that may take on no more than $(m+1)^{\alpha s-1}$ 
different values,\footnote{
To see why this is true, observe that the sum that defines $\bl_i$
depends on $\bx_i=x_{(i-1)m+1}^{im}$ and
$\bz_i=z_{(i-1)m+1}^{im}$ only via the joint type class of pairs 
$(x,z)\in\calX\times\calZ$, associated with $(\bx_i,\bz_i)$.
Thus, the number of different values that
$\bl_i$ may take cannot exceed the total number of such type classes, which
in turn is upper bounded by
$(m+1)^{\alpha s-1}$.}
and hence the same is true concerning the random variable $L$, and so,
$H(L)\le(\alpha s-1)\log(m+1)$. Thus,
\begin{eqnarray}
\label{hk}
\sigma_E(x^n)&=&\frac{1}{m}H(K^{m}|L)\nonumber\\
&=&\frac{1}{m}[H(K^{m})-I(K^{m};L)]\nonumber\\
&\ge&\frac{1}{m}[H(K^{m})-H(L)]\nonumber\\
&\ge&\frac{1}{m}[H(K^{m})-(\alpha s-1)\log(m+1)].
\end{eqnarray}
Now, for all large $m$,
\begin{eqnarray}
\label{central}
H(K^{m})&\ge&H(K^{m}|Y^{m})\nonumber\\
&\ge& I(K^{m};X^{m}|Y^{m})\nonumber\\
&=& H(X^{m}|Y^{m})-H(X^{m}|Y^{m},K^{m})\nonumber\\
&=& H(X^{m})-H(X^{m}|Y^{m},K^{m})\nonumber\\
&\ge&
H(X^{m})-H(X^{m}|Y^{m},K^{m},Z,Z')-
I(Z,Z';X^{m}|Y^{m},K^{m})\nonumber\\
&=&H(X^{m})-0-I(Z,Z';X^{m}|Y^{m},K^{m})\nonumber\\
&\ge&H(X^{m})-H(Z,Z'|Y^{m},K^{m})\nonumber\\
&\ge&H(X^{m})-2\log s,
\end{eqnarray}
where the second equality is due to the 
perfect security assumption and the third
equality is due to the IL property, assuming that $m$ 
is sufficiently large. Thus, combining eqs.\ (\ref{hk}) and (\ref{central}), we obtain
\begin{equation}
\label{empent}
\sigma_E(x^n)\ge \frac{H(X^{m})}{m}-\frac{2\log
s}{m}-(\alpha s-1)\cdot\frac{\log(m+1)}{m}.
\end{equation}
Now, the main term, $H(X^{m})/m$, 
is nothing but the normalized $m$--th order empirical entropy
associated with $x^n$. 
Next, as discussed earlier, we
further lower bound $H(X^m)/m$ in terms of $\rho_{LZ}(x^n)$ at the (small) price of
reducing the bound further by additional terms that will be shown later to be
negligible. In particular, in the sequel, we prove the following inequality:
\begin{equation}
\label{clogclb}
\frac{H(X^{m})}{m}\ge \frac{c(x^n)\log c(x^n)}{n}-
\frac{2m(\log\alpha+1)^2}{(1-\epsilon_n)\log n}-
\frac{2m\alpha^{2m}\log\alpha}{n}-\frac{1}{m}.
\end{equation}
where $\epsilon_n\to 0$ as $n\to \infty$.
Combining this with eq.\ (\ref{empent}), we get
\begin{equation}
\sigma_E(x^n)\ge \frac{c(x^n)\log c(x^n)}{n}-\delta_s(n,m)
\end{equation}
where
\begin{equation}
\delta_s(n,m)=\frac{2\log
s}{m}+(\alpha s-1)\cdot\frac{\log(m+1)}{m}+
\frac{2m(\log\alpha+1)^2}{(1-\epsilon_n)\log n}+
\frac{2m\alpha^{2m}\log\alpha}{n}+\frac{1}{m}.
\end{equation}
We now have the freedom to let $m=m_n$ grow slowly enough
as a function of $n$ such that $\delta_s(n)=\delta_s(n,m_n)$
will vanish for every fixed $s$.
By letting $m_n$ be proportional to $\sqrt{\log n}$, 
$\delta_s(n)$ becomes $O(s\log(\log n)/\sqrt{\log n})$.
Note that the first two terms of $\delta_s(n,m)$ come from
considerations pertaining to encryption, whereas the other terms
appear also in compression. The second term turns out to be the
dominant one, which means that in the encryption problem we end up
with slower decay of the redundancy.  If we compare the difference between
the upper bound and the lower bound in compression (coding them and converse
in \cite{ZL78}), this difference is dominated by a term that is
$O(([\log(2\alpha)]\log(8\alpha s^2)/\log n)$, whereas in encryption the difference
is $O(\alpha s\log(\log n)/\sqrt{\log n})$, namely, a significantly slower
decay rate.

It remains then to establish eq.\ (\ref{clogclb}). To this end, 
let us first recall the analogous setup of lossless compression of individual
sequences using finite--state machines \cite{ZL78}.
A $q$-state
encoder $C$ is defined by a quintuplet
$(\Sigma,\calB,\calX,f,g)$, where
$\Sigma$ is the state set of size $q$,
$\calB$ is
a finite set of binary words
(possibly of different lengths, including the null word for idling),
$\calX$ is the finite
alphabet of the source to be compressed,
$f:\Sigma\times\calY\to\calB$ is the encoder output function, and
$g:\Sigma\times\calY\to\Sigma$ is the next--state function.
When an input sequence $(x_1,x_2,...)$ 
is fed sequentially into $C=(\Sigma,\calX,\calB,f,g)$, the
encoder outputs a sequence of binary words $(b_1,b_2,...)$, $b_i\in\calB$,
while going through a sequence of states $(\sigma_1,\sigma_2,...)$,
according to
\begin{equation}
\label{fsm}
b_i=f(\sigma_i,x_i),~~~\sigma_{i+1}=g(\sigma_i,x_i),~~~i=1,2,...
\end{equation}
where $\sigma_i$ is the state of $C$ at time instant $i$.
A finite--state encoder $C$
is said to be {\it information
lossless} (IL) if for all $\sigma_i\in\Sigma$ and all
$x_i^{i+j-1}\in\calX^n$, $j\ge 1$,
the triple $(\sigma_i,\sigma_{i+j},\bb)$ uniquely determines $x_i^{i+j-1}$,
where $\sigma_{i+j}$ and $\bb=(b_i,...,b_{i+j-1})$ are obtained by iterating
eq.\ (\ref{fsm}) with initial state $\sigma_i$
and $x_i^{i+j-1}$ as input. The length function
associated with $C$ is defined as $\ell_C(x^n)=\sum_{i=1}^n\ell(b_i)$,
where $\ell(b_i)$ is the length of the binary string $b_i\in\calB$.

Consider the incremental parsing of $x^n$
and let $c(x^n)$ be defined
as above. According to \cite[Theorem 1]{ZL78},
for any $q$-state IL encoder
and for every $x^n\in\calX^n$, $n \ge 1$,
\begin{equation}
\label{lowerbound}
\ell_C(x^n) \ge
[c(x^n)+q^2]\log \frac{c(x^n)}{4q^2} .
\end{equation}
Consider next the
Shannon code, operating on $x^n$ by successively encoding its $m$--blocks,
$\bx_1$, $\bx_2$,$\ldots$, $\bx_{n/m}$,
using an arbitrary
probability distribution $Q$. According to this code, $\bx_i$
is encoded
using $\lceil-\log Q(\bx_i)\rceil$ bits,
and so, its length function is given by
\begin{eqnarray}
\ell_S(x^n)&=&\sum_{i=1}^{n/m} \lceil-\log
Q(\bx_i)\rceil\nonumber\\
&=&\frac{n}{m}\sum_{a^{m}}P_{X^{m}}(a^{m})\lceil-\log
Q(a^{m})\rceil\nonumber\\
&\le&\frac{n}{m}\sum_{a^{m}}P_{X^{m}}(a^{m})[-\log Q(a^{m})+1]\nonumber\\
&=&-\frac{n}{m}\sum_{a^{m}}P_{X^{m}}(a^{m})\log
Q(a^{m})+\frac{n}{m}.
\end{eqnarray}
It is easy to see that this code can be implemented by a finite--state encoder
in the following manner: At the beginning of each block ($t~\mbox{mod}~m
=1$), the encoder
is always at some fixed initial state $\sigma_0$. At time
instant $t=(i-1)m+j$, $1< j \le m$, the state
$\sigma_t$ is defined as $x_{(i-1)m+1}^{(i-1)m+j-1}$.
The encoder outputs the null
string whenever $t~\mbox{mod}~m\ne 1$; when $t~\mbox{mod}~m =1$, 
the encoder emits the Shannon codeword
of the block just terminated. The total number of states
is therefore $q=\sum_{j=0}^{m-1}\alpha^j 
=(\alpha^{m}-1)/(\alpha-1)$.
It is also easy to see that the Shannon code is IL. For given positive integers
$i$ and $j$, suppose we are given $\sigma_i$, $\sigma_{i+j}$,
and $(b_i,b_{i+1},...,b_{i+j-1})$. Then
$(x_i,x_{i+1},...,x_{i+j-1})$ can be reconstructed as follows.
If time instants $i$ and $i+j$ fall in the same $m$-block then
$\sigma_{j+1}$ conveys full information on
$(x_i,x_{i+1},...,x_{i+j-1})$. Otherwise, we use the following procedure:
The segment from time $i$ until
the end of the current block is reconstructed by decoding the codeword emitted
at the end of this block.
Similarly, if there are any additional blocks that are
fully contained in the segment from $i$ to $i+j$, they can also be
reconstructed by decoding. Finally, the portion of the last
block until position $i+j-1$ can be recovered again from the final state.

It now follows that the length function of the Shannon code must satisfy the lower
bound (\ref{lowerbound}) with $q=q_m\dfn(\alpha^{m}-1)/(\alpha-1)\le\alpha^{m}$, and so,
\begin{equation}
-\frac{n}{m}\sum_{a^{m}}P_{X^{m}}(a^{m})\log
Q(a^{m})+\frac{n}{m} \ge
[c(x^n)+q_m^2]
\log \frac{c(x^n)}{4q_m^2}.
\end{equation}
Since this holds for every $Q$ while the right--hand side is independent
of $Q$, we may minimize the left--hand side w.r.t.\ $Q$ and obtain
\begin{eqnarray}
\frac{n}{m}H(X^{m})+\frac{n}{m}
&\ge&[c(x^n)+q_m^2]
\log \frac{c(x^n)}{4q_m^2}\nonumber\\
&\ge&c(x^n)\log c(x^n)-c(x^n)\log(4q_m^2)-q_m^2\log(4q_m^2)\nonumber\\
&\ge&c(x^n)\log c(x^n)-c(x^n)\log(4\alpha^{2m})-\alpha^{2m}\log(4\alpha^{2m})\nonumber\\
&\ge&c(x^n)\log c(x^n)-2m c(x^n)(1+\log\alpha)-2m\alpha^{2m}(1+\log\alpha)\nonumber\\
&\ge&c(x^n)\log c(x^n)-\frac{2m n (1+\log\alpha)^2}{(1-\epsilon_n)\log n}-
2m\alpha^{2m}(1+\log\alpha),
\end{eqnarray}
where the last inequality follows from \cite[eq.\ (6)]{ZL78}.
Eq.\ (\ref{clogclb}) is now obtained by normalizing both sides by $n$.
This completes the proof of Theorem 1.\\

\section{Extensions}

In this section, we extend Theorem 1 in two directions, availability of SI
and lossy reconstruction. As described in the Introduction, we first
consider each one of these directions separately, and then jointly.

\subsection{Availability of Side Information}
\label{si}

Consider the case where SI is available
at the encrypter/decrypter/eavesdropper.
Suppose that, in
addition to the source sequence $\bx$, there is an (individual) SI sequence
$\bs=(s_1,s_2,\ldots)$, $s_i\in\calS$, $i=1,2,\ldots$, where $\calS$ is a
finite alphabet. Let us assume first that all three parties (encoder, decoder, and
eavesdropper) have access to $\bs$. In the formal model definition, a few
modifications are needed: 
\begin{enumerate}
\item In eqs.\ (\ref{ti}), (\ref{yi}), and (\ref{nextstate}), the functions $\Delta$,
$f$ and $g$ should be allowed to depend on the additional argument $s_i$, 
\item The
definition of perfect security should allow conditioning on
$\bs$, in addition to the present conditioning on $\bx$. I.e.,
$\mbox{Pr}\{Y_n^m=y_n^m|\bx,\bs\}$ is independent of $\bx$ 
for all positive integers $n$, $m$ (but it is allowed to depend
on $\bs$).
\item In the definition of an IL encrypter, the quadruple
$(z,k^n,f(z,x^n,k^n),g(z,x^n))$ should be extended to be the quintuple
$(z,k^n,s^n,f(z,x^n,k^n),g(z,x^n))$.
\end{enumerate}

In Theorem 1, the LZ complexity of $x^n$,
should be replaced by the conditional
LZ complexity of $x^n$ given $s^n$, denoted $\rho_{LZ}(x^n|s^n)$, which is an empirical measure of
conditional entropy (or conditional compressibility), that is defined as
follows (see also \cite{Merhav00}, \cite{Ziv85}):
Given $x^n$ and $s^n$,
let us apply the incremental
parsing procedure of the LZ algorithm
to the sequence of pairs $((x_1,s_1),(x_2,s_2),\ldots,(x_n,s_n))$.
According to this procedure, all phrases are distinct
with a possible exception of the last phrase, which might be incomplete.
Let $c(x^n,s^n)$ denote the number of distinct phrases.
For example,\footnote{The same example appears in \cite{Ziv85}.} if
\begin{eqnarray}
x^6&=&0~|~1~|~0~0~|~0~1|\nonumber\\
s^6&=&0~|~1~|~0~1~|~0~1|\nonumber
\end{eqnarray}
then $c(x^6,s^6)=4$.
Let $c(s^n)$ denote the resulting number of distinct phrases
of $s^n$, and let $\bs(l)$ denote the $l$th distinct $\bs$--phrase,
$l=1,2,...,c(s^n)$. In the above example, $c(s^6)=3$. Denote by
$c_l(x^n|s^n)$ the number of occurrences of $\bs(l)$ in the
parsing of $s^n$, or equivalently, the number of distinct $\bx$-phrases
that jointly appear with $\bs(l)$. Clearly, $\sum_{l=1}^{c(s^n)} c_l(x^n|s^n)=
c(x^n,s^n)$. In the above example, $\bs(1)=0$, $\bs(2)=1$, $\bs(3)=01$,
$c_1(x^6|s^6)=c_2(x^6|s^6)=1$, and $c_3(x^6|s^6)=2$. Now, the conditional LZ
complexity of $x^n$ given $s^n$ is defined as
\begin{equation}
\rho_{LZ}(x^n|s^n)=\frac{1}{n}\sum_{l=1}^{c(s^n)}c_l(x^n|s^n)\log c_l(x^n|s^n).
\end{equation}
The proof of Theorem 1 extends quite straightforwardly: The definition
of $P_{K^{m}X^{m}Y^{m}ZZ'L}$ should be extended to
$P_{K^{m}S^{m}X^{m}Y^{m}ZZ'L}$ in account of the
empirical distribution that includes the $m$--blocks of $s^n$.
In (\ref{central}), all the conditionings should include $S^{m}$ 
in addition to all existing conditionings, resulting in the inequality
\begin{equation}
H(K^{m})\ge H(X^{m}|S^{m})-2\log s. 
\end{equation}
Finally, $H(X^{m}|S^{m})$
is further lower bounded in terms of $\rho_{LZ}(x^n|s^n)$ since the latter is
essentially a lower bound on the the compression ratio of $x^n$ given $s^n$
using finite--state encoders (see \cite[eq.\ (13)]{Merhav00}).
The direct is obtained by first, compressing $x^n$
to about $n\cdot \rho_{LZ}(x^n|s^n)$ bits using the conditional parsing scheme
\cite[Lemma 2, eq.\ (A.11)]{Ziv85} and then applying one--time pad encryption. 

The same
performance can be achieved even if the encrypter does not have access to
$s^n$, by using a scheme in the spirit 
of Slepian--Wolf coding: Randomly assign to each member
of $\calX^n$ a bin, selected independently at random across the set
$\{1,2,\ldots,2^{nR}\}$. The encrypter applies one--time pad to the 
$(nR)$--bit binary
representation of the bin index of $x^n$. The decrypter, first decrypts the
bin index using the key and then seeks a sequence $\hat{x}^n$ within the
given bin, which satisfies $\rho_{LZ}(\hat{x}^n|s^n) < R-\epsilon$. If there is one
and only one such sequence, then it becomes the decoded message, otherwise an
error is declared. This scheme works, just like the ordinary
SW coding scheme, because the number of $\{\hat{x}^n\}$
for which $\rho_{LZ}(\hat{x}^n|s^n) < R-\epsilon$
does not exceed $2^{n[R-\epsilon+O(\log(\log n)/\log n)]}$ \cite[Lemma 2]{Ziv85}. 
The weakness of this is that prior
knowledge of (a tight upper bound on) $\rho_{LZ}(x^n|s^n)$ is required. If, for example, it is known
that $x^n$ is a
noisy version of $s^n$, generated, say, by a known additive channel, 
then $R$ should be 
essentially the entropy rate of the noise. 

The case where the legitimate receiver and the eavesdropper have access to
different SI's will be discussed in Subsection \ref{silossy}, where we also extend the
scope to lossy reconstruction.

\subsection{Lossy Reconstruction}
\label{lossy}

Suppose that we are content with
a lossy reconstruction, $\hat{x}^n$, at the legitimate receiver.
In general, this reconstruction may be a random vector
due to possible dependence on the random key bits.
It is required, however, that 
$d(x^n,\hat{x}^n)\le nD$ with probability one, for some distortion measure $d$.
Then, in Theorem 1, $\rho_{LZ}(x^n)$ should be replaced by
the ``LZ rate--distortion function'' of $x^n$, which is defined as
\begin{equation}
r_{LZ}(D;x^n)\dfn \min_{\{\hat{x}^n:~d(x^n,\hat{x}^n)\le nD\}} \rho_{LZ}(\hat{x}^n).
\end{equation}
In the proof of Theorem 1, the joint distribution 
$P_{K^{m}X^{m}\hat{X}^{m}Y^{m}ZZ'L}$ should be defined as
the expectation (w.r.t.\ the randomness of the key)
of the $m$--th order empirical distribution extracted
from the sequences $(k^n,x^n,\hat{x}^n,y^n)$ and the resulting states
$\{z_{(i-1)m+1}\}_{i=1}^{n/m}$ and key lengths
$\{\bl_i\}_{i=1}^{n/m}$.
The definition
of the IL property can be 
slightly relaxed to a notion of ``nearly IL'' (NIL) property, which
allows recovery with small uncertainty for all large enough $n$.
In particular, we shall assume that given 
$w\dfn (z_i,k_i^{i+n},f(z_i,x_i^{n+i-1},k_i^{n+i-1}),g(z_i,x_i^{n+i-1}))$,
$\hat{x}_i^{n+i-1}$ must lie, with probability one, in a subset 
$\calA_n(w)\subset
\hat{\cal X}^n$,
where\footnote{
This might be the case if unambiguous reconstruction of
$\hat{x}_i^{n+i-1}$ requires additional information from times later than
$t=n+i-1$. For example, if the encrypter works in
blocks of fixed size $m$, $\hat{x}^n$ is deterministic, 
and $n \gg m$, then by viewing the block code as finite--state machine as
before, there might be uncertainty 
in not more than the $m$ last 
symbols of $\hat{x}^n$ in case the 
last block is incomplete (e.g., when $m$ does not 
divide $n$ or the $n$--block 
considered is not synchronized to the $m$--blocks).
In this case, 
$|\calA_n(w)|\le |\hat{\calX}|^m$, which is fixed,
independent of $n$, and so $\eta_n=O(1/n)$.} 
\begin{equation}
\eta_n\dfn\lim_{n\to\infty}
\frac{1}{n}\log\max_w|\calA_n(w)|= 0.
\end{equation}
Perfect security should be defined as statistical independence between
the cryptogram and both the source and reconstruction, i.e., the
probability of any segment of $\{y_i\}$ should not depend on
either $\bx$ or $\hat{\bx}$.

In the proof of the converse part, in eq.\ (\ref{central}),
$X^{m}$ should be replaced by $\hat{X}^{m}$ in all places, and
we get 
\begin{equation}
H(K^{m})\ge H(\hat{X}^{m})-2\log s-m\eta_m, 
\end{equation}
as $H(\hat{X}^{m}|Y^{m},K^{m},Z,Z')/m$
would be upper bounded by $\eta_{m}$.
Then, $H(\hat{X}^{m})/m$ is
further lower bounded in terms of $\bE \rho_{LZ}(\hat{x}^n)$, essentially in the
same way as before, where
here we have also used the fact that,
due to the concavity of the entropy functional, 
$H(\hat{X}^{m})$ is lower bounded by
the expected $m$--th order 
conditional empirical entropy pertaining to the realizations of
$\hat{x}^n$.
Finally, since we require $d(x^n,\hat{x}^n)\le nD$ with probability one,
then $\bE \rho_{LZ}(\hat{x}^n)$ is trivially further lower bounded by $r_{LZ}(D;x^n)$.

Again, the direct is obvious, and it implies that 
at least asymptotically, there is nothing to gain
from randomizing the reconstruction: The best choice of $\hat{x}^n$ is the
one with minimum LZ complexity within the sphere of radius $nD$ around $x^n$.
This conclusion is not obvious a--priori as one might speculate that
a randomized reconstruction, depending on the key, 
may potentially be more secure than a
deterministic one.

Note that we have not assumed anything on the distortion measure $d$, not
even additivity. Another difference between Theorem 1 of the lossless case
and its present extension to the lossy case, is that we are know longer
able to characterize the rate of convergence of $\delta_s(n)$, as it depends on
the rate of decay of $\eta_{m}$. In fact,
we could have replaced the
IL property we assumed in the lossless case 
by the NIL property there too, but again,
the cost would be the loss the ability to specify the behavior of $\delta_n$.

\subsection{Lossy Reconstruction With Side Information}
\label{silossy}

The simultaneous extension of Theorem 1, allowing both distortion $D$
and SI $s^n$ leads, with the obvious modifications, to
$\min\{\rho_{LZ}(\hat{x}^n|s^n):~d(x^n,\hat{x}^n)\le nD\}$, whose achievability
is conceptually straightforward when all parties 
have access to $s^n$, including the encrypter.
But what if the encrypter does not have access to $s^n$?

In this case, there is no longer an apparent way to characterize the
minimum key rate that must be consumed in terms of LZ complexities.
This should not be surprising in view of the fact that
even in the less involved problem of pure lossy compression of individual
sequences with SI available 
at the decoder, performance is no longer characterized in terms of the LZ
complexity (see, e.g., \cite{MZ06} and references therein). Similarly as in
\cite{MZ06}, here we are able
to give a certain characterization for the case where the decrypter is also
modeled as an FSM. While our performance characterization may not seem very
explicit, the main message behind it (like in \cite{MZ06}) is that the
performance of the best $s$--state encrypter--decrypter can be achieved by
block codes of length $m$ within a redundancy term that decays as $m\to
\infty$ for every fixed $s$.

Referring to our definition of the finite--state encrypter in
Section 2, we 
model the decrypter as a device that implements the
following recursive equations:
\begin{eqnarray}
z_{i+1}'&=&g'(z_i',y_i,s_i)~~~~i=1,2,\ldots\\
\hat{x}_{i-\tau}&=&f'(z_i',y_i,s_i,k_i)~~~i=\tau+1,\tau+2,\ldots
\end{eqnarray}
where $\tau$ (non--negative integer) is the encoding--decoding delay and
$z_i'\in\calZ$
is the state of the decrypter at time $t$.
We also model the channel from $x^n$ to $s^n$ as a memoryless channel
\begin{equation}
P_{S^n|X^n}(s^n|x^n)=\prod_{i=1}^nP_{S|X}(s_i|x_i).
\end{equation}

We argue that the minimum key rate consumed by any finite--state
encrypter--decrypter with $s$ states and delay $\tau$ is lower
bounded by $r_s(D+\tau
d_{\max}/m)$,
where $d_{\max}=\max_{x,\hx}d(x,\hx)$ is assumed finite and where
$r_s(D)$ is defined as
the minimum of $H(K^m|L)/m=\frac{1}{m}\sum_{l\ge 1} l\cdot P_L(l)$
over all random variables $(\hY,W,L)$ such that: (i) the support of
$P_L$ is of size $(m+1)^{\alpha s-1}$, (ii) $X^m\to S^m\to \hY$ is a Markov
chain (perfect security), (iii) $\min_h\bE\{d(X^m,h(W,L,U^L,S^m,\hY))\le
mD$, (iv) $(W,L)\to X^m\to S^m$ is a Markov chain and $\hY=g(W,L,X^m,U^L)$
for some deterministic function $g$, (v) the alphabet size of
$W$ is $s^2$, and (vi) the alphabet size of $\hY$ is the minimum needed (by
the Carath\'eodory theorem) in order to maintain (i)-(v).

Consider again the partition of a block of length $n$ into $n/m$
non--overlapping blocks, each of length $m$, along with the induced joint empirical distribution
$P_{K^mX^mY^mZZ'L}$ defined as before except that now $Z'$ is the random
variable that designates the relative 
frequency of the state of the decrypter $z_t'$ at times $t=im+1$,
$i=1,2,\ldots,n/m$.
Next, define
\begin{equation}
\label{jointpmf}
P_{K^mS^mX^mY^mZZ'L}=P_{K^mX^mY^mZZ'L}\times P_{S^m|X^m}.
\end{equation}
First, observe that 
$y_{im+1}^{im+m}$ depends (deterministically) only on $z_{im+1}$,
$x_{im+1}^{im+m}$, and $k_{im+1}^{im+m}$. Similarly
$\hx_{im-\tau+1}^{im-\tau+m}$ depends only on
$k_{im+1}^{im+m}$, $s_{im+1}^{im+m}$, $y_{im+1}^{im+m}$, $z_{im+1}'$ and $l$.
Let us denote then
$$y_{im-\tau+1}^{im-\tau+m}=
g(z_{im+1},l,x_{im+1}^{im+m},k_{im+1}^{im+m})$$
and
$$\hx_{im-\tau+1}^{im-\tau+m}=
h(z_{im+1}',l,k_{im+1}^{im+m},s_{im+1}^{im+m},y_{im+1}^{im+m}),$$
Assuming that $m > \tau$,
let $\hx_{im+1}^{im-\tau+m}\dfn 
h'(z_{im+1}',l,k_{im+1}^{im+m},s_{im+1}^{im+m},y_{im+1}^{im+m})$,
be defined simply by truncating the first $\tau$ components of
$h(z_{im+1}',l,k_{im+1}^{im+m},s_{im+1}^{im+m},y_{im+1}^{im+m})$.
Next, extend the definition of $P_{K^mS^mX^mY^mZZ'L}$ to 
$P_{K^mLX^m\hX^{m-d}Y^mS^mZ}$ by defining $P_{\hX^{m-\tau}|K^mLS^mX^mY^mZ'L}$ as
a degenerate PMF that puts all its mass on
$\hX^{m-\tau}=h'(Z',L,K^m,S^m,Y^m)=h'(Z',L,U^L,S^m,Y^m)$. 
Now observe that eq.\ (\ref{jointpmf}) implies that
$(Y^m,Z,Z',L)\to X^m\to S^m$ is a Markov chain.
For the purpose of obtaining a lower bound 
on the performance of finite--state encryption--decryption systems
on the consumed key rate, it
is legitimate to let $g$ include dependence on $Z'$ and to let 
$h$ include dependence on $Z$ in addition to their dependencies on 
the other variables involved. By doing so, the random variables $Z$ and
$Z'$ appear together in all relevant places of the characterization and thus,
we can define $W=(Z,Z')$ which is a random variable whose alphabet size is
$s^2$. The (variable--length) string $Y^m$ can be replaced by a single random
variable $\hY$ with the suitable alphabet size as defined above.

As for the distortion, we have
\begin{eqnarray}
D&\ge&\frac{1}{n}\bE\{d(x^n,\hat{X}^n)\}\nonumber\\
&\ge&\frac{1}{m}\bE\{d(X^{m-\tau},\hat{X}^{m-\tau})\}\nonumber\\
&\ge&\frac{1}{m}[\bE\{d(X^m,\hat{X}^m)\}-\tau\cdot d_{\max}]
\end{eqnarray}
where $\hX^m$ is defined by concatenating $\hX^{m-\tau}$ with a
random $\tau$--vector in $\hat{\calX}^\tau$
that is an arbitrary function of $(Z',L,U^L,S^m,Y^m)$ (or $(W,L,U^L,S^m,Y^m)$). Thus, the minimum
required key rate, $H(K^m|L)/m$, of any $s$--state encrypter--decrypter cannot be smaller than
$r_s(D+\tau\cdot d_{\max}/m)$ by definition.

We can achieve this performance by block codes as follows. 
For a given empirical distribution of $X^m$ and SI channel $P_{S|X}$, find the optimum distribution
conditional distribution $P_{WL|X^m}$, the encrypter $g$ and the decrypter 
$h$ that achieve $r_s(D)$. For every
$\bx_i$, $i=1,2,\ldots,n/m$, apply the channel $P_{WL|X^m}$ to 
generate $w_i$ and $l_i$ given $\bx_i$,
and then compute $\hy_i=g(w_i,l_i,\bx_i,\bu^{l_i})$.
Next, transmit  
$\hy_i$ plus one--time pad encrypted versions of $w_i$ and $l_i$ (to avoid any
leakage of information concerning $\bx_i$ via these random variables).
These encryptions
of $w_i$ and $l_i$ require extra key rates of $(2\log s)/m$ and $(\alpha
s-1)(\log m)/m$, respectively. The information concerning 
the optimum $h$ should be transmitted to the decrypter once in an $n$--block.
Its one--time pad encryption requires additional key rate given by the
description length of $h$, which depends only on $m$ (as well as the alphabet
sizes), normalized by $n$, and hence
it is negligible when $n \gg m$. The decrypter simply applies the decoding
function $h$ and outputs the reconstruction.

Finally, note that if the eavesdropper has another version of the side
information sequence, say, $\tilde{s}^n$ (generated from $x^n$ by another known
memoryless channel $P_{\tilde{S}|X}$), everything remains the same except
that the perfect security requirement (ii) is replaced by $X^m\to \tilde{S}^m
\to \hY$.

\end{document}